\documentclass{amsart}
\usepackage{amsfonts}
\usepackage{amscd}

\input{tcilatex}
\theoremstyle{definition}
\theoremstyle{remark}
\numberwithin{equation}{section}

\input tcilatex

\begin{document}
\title{On Quantum It\^{o} Algebras\\
and their decompositions.}
\author{V. P. Belavkin.}
\address{Mathematics Department, University of Nottingham,\\
NG7 2RD, UK.}
\email{vpb@maths.nott.ac.uk}
\date{December 20, 1997}
\subjclass{Quantum Probability and Stochastic Analysis}
\thanks{}
\maketitle

\begin{abstract}
A simple axiomatic characterization of the noncommutative It\^{o} algebra is
given and a pseudo-Euclidean fundamental representation for such algebra is
described. It is proved that every quotient It\^{o} algebra has a faithful
representation in a Minkowski space and is canonically decomposed into the
orthogonal sum of quantum Brownian (Wiener) algebra and quantum L\'{e}vy
(Poisson) algebra. In particular, every quantum thermal noise of a finite
number of degrees of freedom is the orthogonal sum of a quantum Wiener noise
and a quantum Poisson noise as it is stated by the L\'{e}vy-Khinchin theorem
in the classical case. Two basic examples of non-commutative It\^{o} finite
group algebras are considered.
\end{abstract}

\section{Introduction}

The classical differential calculus for the infinitesimal increments $%
\mathrm{d}x=x\left( t+\mathrm{d}t\right) -x\left( t\right) $ became
generally accepted only after Newton gave a very simple algebraic rule $%
\left( \mathrm{d}t\right) ^{2}=0$ for the formal computations of first order
differentials for smooth trajectories $t\mapsto x\left( t\right) $ in a
phase space. The linear space of the differentials $\mathrm{d}x=\alpha 
\mathrm{d}t$ for a (complex) trajectory became treated at each $x=x\left(
t\right) \in \mathbb{C}$ as a one-dimensional algebra $\mathfrak{a}=\mathbb{C%
}d_{t}$ of the elements $a=\alpha d_{t}$ with involution $a^{\star }=\bar{%
\alpha}d_{t}$ given by the complex conjugation $\alpha \mapsto \bar{\alpha}$
of the derivative $\alpha =\mathrm{d}x/\mathrm{d}t\in \mathbb{C}$ and the
nilpotent multiplication $a\cdot a^{\star }=0$ corresponding to the
multiplication table $d_{t}\cdot d_{t}^{\star }=0$ for the basic nilpotent
element $d_{t}=d_{t}^{\star }$, the abstract notation of $\mathrm{d}t$. Note
that the nilpotent $\star $-algebra $\mathfrak{a}$ of abstract
infinitesimals $\alpha d_{t}$ has no realization in complex numbers, as well
as no operator representation $\alpha D_{t}$ with a Hermitian nilpotent $%
D_{t}=D_{t}^{\dagger }$ in a Euclidean (complex pre-Hilbert) space, but it
can be represented by the algebra of complex nilpotent $2\times 2$ matrices $%
\hat{a}=\alpha \hat{d}_{t}$, where $\hat{d}_{t}=\frac{1}{2}\left( \hat{\sigma%
}_{3}+i\hat{\sigma}_{1}\right) =\hat{d}_{t}^{\dagger }$ with respect to the
standard Minkowski metric $\left( \mathrm{x}|\mathrm{x}\right) =\left\vert
\zeta \right\vert ^{2}-\left\vert \eta \right\vert ^{2}$ for $\mathrm{x}%
=\zeta \mathrm{e}_{+}+\eta \mathrm{e}_{-}$ in $\mathbb{C}^{2}$. The complex
pseudo-Hermitian nilpotent matrix $\hat{d}_{t}$, $\hat{d}_{t}^{2}=0$,
representing the multiplication $d_{t}^{2}=d_{t}\cdot d_{t}=0$, has the
canonical triangular form 
\begin{equation}
\text{$\mathrm{D}_{t}$}=\left[ 
\begin{array}{ll}
0 & 1 \\ 
0 & 0%
\end{array}%
\right] ,\quad \text{$\mathrm{D}_{t}$}^{\dagger }=\left[ 
\begin{array}{ll}
0 & 1 \\ 
1 & 0%
\end{array}%
\right] \text{$\mathrm{D}_{t}$}^{\ast }\left[ 
\begin{array}{ll}
0 & 1 \\ 
1 & 0%
\end{array}%
\right] =\text{$\mathrm{D}_{t}$},\quad \text{$\mathrm{D}_{t}$}^{\ast }=\left[
\begin{array}{ll}
0 & 0 \\ 
1 & 0%
\end{array}%
\right] \quad   \label{1}
\end{equation}%
in the basis $\mathrm{k}_{\pm }=\left( \mathrm{e}_{+}\pm \mathrm{e}%
_{-}\right) /\sqrt{2}$ in which $\left( \mathbf{x}\text{$\mathrm{D}_{t}$}|%
\mathbf{x}\right) =\overline{\left( \mathbf{x}\text{$\mathrm{D}_{t}$}|%
\mathbf{x}\right) }$ for all $\mathbf{x}=\left( \xi _{-},\xi _{+}\right) $
with respect to the pseudo-Euclidean scalar product $\left( \mathbf{x}|%
\mathbf{x}\right) =\xi _{-}\xi ^{-}+\xi _{+}\xi ^{+}$, where $\xi ^{\pm
}=\left( \zeta \pm \eta \right) /\sqrt{2}=\bar{\xi}_{\mp }\in \mathbb{C}$.

The Newton's formal computations can be generalized to non-smooth paths to
include the calculus of first order forward differentials $\mathrm{d}y\simeq
\left( \mathrm{d}t\right) ^{1/2}$ of continuous diffusions $y\left( t\right)
\in \mathbb{R}$ which have no derivative at any $t$, and the forward
differentials $\mathrm{d}n\in \left\{ 0,1\right\} $ of left continuous
counting trajectories $n\left( t\right) \in \mathbb{Z}_{+}$ which have zero
derivative for almost all $t$ (except the points of discontinuity when $%
\mathrm{d}n=1$). The first is usually done by adding the rules 
\begin{equation}
\left( \mathrm{d}w\right) ^2=\mathrm{d}t,\quad \mathrm{d}w\mathrm{d}t=0=%
\mathrm{d}t\mathrm{d}w  \label{2}
\end{equation}
in formal computations of continuous trajectories having the first order
forward differentials $\mathrm{d}x=\alpha \mathrm{d}t+\beta \mathrm{d}w$
with the diffusive part given by the increments of standard Brownian paths $%
w\left( t\right) $. The second can be done by adding the rules 
\begin{equation}
\left( \mathrm{d}m\right) ^2=\mathrm{d}m+\mathrm{d}t,\quad \mathrm{d}m%
\mathrm{d}t=0=\mathrm{d}t\mathrm{d}m  \label{3}
\end{equation}
in formal computations of left continuous and smooth for almost all $t$
trajectories having the forward differentials $\mathrm{d}x=\alpha \mathrm{d}%
t+\gamma \mathrm{d}m$ with jumping part $\mathrm{d}z\in \left\{ \gamma
,-\gamma \mathrm{d}t\right\} $ given by the increments of standard L\'{e}vy
paths $m\left( t\right) =n\left( t\right) -t$. These rules, well known since
the beginning of this century, were formalized by It\^{o} \cite{Ito51} into
the form of a stochastic calculus: the first one is now known as the
multiplication rule for the forward differential of the standard Wiener
process $w\left( t\right) $, and the second one is the multiplication rule
for the forward differential of the standard Poisson process $n\left(
t\right) $, compensated by its mean value $t$.

The linear span of $\mathrm{d}t$ and $\mathrm{d}w$ forms a two-dimensional
differential It\^{o} algebra $\mathfrak{b}=\mathbb{C}d_{t}+\mathbb{C}d_{w}$
for the complex Brownian motions $x\left( t\right) =\int \alpha \mathrm{d}%
t+\int \eta \mathrm{d}w$, where $d_{w}=d_{w}^{\star }$ is a nilpotent of
second order element, representing the real increment $\mathrm{d}w$, with
multiplication table $d_{w}^{2}=d_{t}$, $d_{w}\cdot d_{t}=0=d_{t}\cdot d_{w}$%
, while the linear span of $\mathrm{d}t$ and $\mathrm{d}m$ forms a
two-dimensional differential It\^{o} algebra $\mathfrak{c}=\mathbb{C}d_{t}+%
\mathbb{C}d_{m}$ for the complex L\'{e}vy motions $x=\int \alpha \mathrm{d}%
t+\int \zeta \mathrm{d}m$, where $d_{m}=d_{m}^{\star }$ is a basic element,
representing the real increment $\mathrm{d}m$, with multiplication table $%
d_{m}^{2}=d_{m}+d_{t}$, $d_{m}\cdot d_{t}=0=d_{t}\cdot d_{m}$. As in the
case of the Newton algebra, the It\^{o} $\star $-algebras $\mathfrak{b}$ and 
$\mathfrak{c}$ have no Euclidean operator realization, but they can be
represented by the algebras of triangular matrices $\mathrm{B}=\alpha 
\mathrm{D}_{t}+\eta \mathrm{D}_{w}$, $\mathrm{C}=\alpha \mathrm{D}_{t}+\zeta 
\mathrm{D}_{m}$ with pseudo-Hermitian basis elements 
\begin{equation}
\text{$\mathrm{D}_{t}$}=\left[ 
\begin{array}{lll}
0 & 0 & 1 \\ 
0 & 0 & 0 \\ 
0 & 0 & 0%
\end{array}%
\right] ,\quad \mathrm{D}_{w}=\left[ 
\begin{array}{lll}
0 & 1 & 0 \\ 
0 & 0 & 1 \\ 
0 & 0 & 0%
\end{array}%
\right] ,\emph{\quad }\mathrm{D}_{m}\mathbf{=}\left[ 
\begin{array}{lll}
0 & 1 & 0 \\ 
0 & 1 & 1 \\ 
0 & 0 & 0%
\end{array}%
\right] ,  \label{4}
\end{equation}%
\begin{equation*}
\text{$\mathrm{D}_{t}$}^{\dagger }=\text{$\mathrm{D}_{t}$},\quad \mathrm{D}%
_{w}^{\dagger }=\mathrm{D}_{w},\quad \mathrm{D}_{m}^{\dagger }=\mathrm{D}_{m}
\end{equation*}%
where $\left( \mathbf{x}\mathrm{B}^{\dagger }|\mathbf{x}\right) =\overline{%
\left( \mathbf{x}\mathrm{B}|\mathbf{x}\right) }$ for all $\mathbf{x}=\left(
\xi _{-},\xi _{\circ },\xi _{+}\right) \in \mathbb{C}^{3}$ in the complex
three-dimensional Minkowski space with respect to the indefinite metric $%
\left( \mathbf{x}|\mathbf{x}\right) =\xi _{-}\xi ^{-}+\xi _{\circ }\xi
^{\circ }+\xi _{+}\xi ^{+}$, where $\xi ^{\mu }=\bar{\xi}_{-\mu }$ with $%
-\left( -,\circ ,+\right) =\left( +,\circ ,-\right) $.

Note that according to the L\'{e}vy-Khinchin theorem, every stochastic
process $x\left( t\right) $ with independent increments can be canonically
decomposed into a smooth, Wiener and Poisson parts as in the mixed case of
one-dimensional complex motion $x\left( t\right) =\int \alpha \mathrm{d}%
t+\int \eta \mathrm{d}w+\int \zeta \mathrm{d}m$ given by the orthogonal and
thus commutative increments $\mathrm{d}w\mathrm{d}m=0=\mathrm{d}m\mathrm{d}w$%
. In fact such generalized commutative differential calculus applies not
only to the stochastic integration with respect to the processes with
independent increments; these formal algebraic rules, or their
multidimensional versions, can be used for formal computations of forward
differentials for any classical trajectories decomposed into the smooth,
diffusive and jumping parts.

Two natural questions arise: are there other then these two commutative
differential algebras which could be useful, in particular, for formal
computations of the noncommutative differentials in quantum theory, and if
there are, is it possible to characterize them by simple axioms and to give
a generalized version of the L\'{e}vy-Khinchin decomposition theorem? The
first question has been already positively answered since the well known
differential realization of the simplest non-commutative table $%
d_{w}d_{w}^{\star }=\rho _{+}d_{t}$, $d_{w}^{\star }d_{w}=\rho _{-}d_{t}$\
for $\rho _{+}>\rho _{-}\geq 0$ was given in the mid of 60-th in terms of
the annihilators $\hat{w}\left( t\right) $ and creators $\hat{w}^{\dagger
}\left( t\right) $ of a quantum Brownian thermal noise \cite{Grd91}. This
paper gives a systematic answer on the second question, the first part of
which has been in principle positively resolved in our papers \cite{3,4}.

Although the orthogonality condition $d_{w}\cdot d_{m}=0=d_{w}\cdot d_{m}$
for the classical independent increments $\mathrm{d}w$ and $\mathrm{d}m$ can
be realized only in a higher, at least four, dimensional Minkowski space, it
is interesting to make sense of the non-commutative $\star $-algebra,
generated by three dimensional non-orthogonal matrix representations (\ref{4}%
) of these differentials with $d_{w}\cdot d_{m}\neq d_{w}\cdot d_{m}$: 
\begin{equation*}
\mathrm{D}_{w}\mathrm{D}_{m}=\left( \mathrm{D}_{m}\mathrm{D}_{w}\right)
^{\dagger }=\left[ 
\begin{array}{lll}
0 & 1 & 1 \\ 
0 & 0 & 0 \\ 
0 & 0 & 0%
\end{array}%
\right] \neq \left[ 
\begin{array}{lll}
0 & 0 & 1 \\ 
0 & 0 & 1 \\ 
0 & 0 & 0%
\end{array}%
\right] =\left( \mathrm{D}_{w}\mathrm{D}_{m}\right) ^{\dagger }=\mathrm{D}%
_{m}\mathrm{D}_{w}.
\end{equation*}%
This is the four-dimensional $\star $-algebra $\mathfrak{a}=\mathbb{C}%
\mathrm{D}_{t}+\mathbb{C}\mathrm{E}_{-}+\mathbb{C}\mathrm{E}^{+}+\mathbb{C}%
\mathrm{E}$ of triangular matrices $\mathrm{A}=\alpha \mathrm{D}+z^{-}%
\mathrm{E}_{-}+z_{+}\mathrm{E}^{+}+z\mathrm{E}$, 
\begin{equation*}
\,\mathrm{E}_{-}=\left[ 
\begin{array}{lll}
0 & 1 & 0 \\ 
0 & 0 & 0 \\ 
0 & 0 & 0%
\end{array}%
\right] ,\;\,\mathrm{E}^{+}\mathbf{=}\left[ 
\begin{array}{lll}
0 & 0 & 0 \\ 
0 & 0 & 1 \\ 
0 & 0 & 0%
\end{array}%
\right] ,\;\,\mathrm{E}=\left[ 
\begin{array}{lll}
0 & 0 & 0 \\ 
0 & 1 & 0 \\ 
0 & 0 & 0%
\end{array}%
\right] ,
\end{equation*}%
where $\mathrm{E}^{+}=\mathrm{GE}_{-}^{\ast }\mathrm{G}=\mathrm{E}%
_{-}^{\dagger }$, $\mathrm{E}=\mathrm{E}^{\dagger }$ with respect to the
Minkowski metric tensor $\mathrm{G}$ in the canonical basis. given by the
algebraic combinations 
\begin{equation*}
\mathrm{E}_{-}=\mathrm{D}_{w}\mathrm{D}_{m}-\mathrm{D}_{t},\quad \mathrm{E}%
^{+}=\mathrm{D}_{m}\mathrm{D}_{w}-\mathrm{D}_{t},\quad \mathrm{E}=\mathrm{D}%
_{m}-\mathrm{D}_{w}
\end{equation*}%
of three matrices (\ref{4}). It realizes the multiplication table 
\begin{eqnarray*}
e_{-}\cdot e^{+} &=&\text{$d_{t}$},\quad e_{-}\cdot e=e_{-}, \\
e\cdot e^{+} &=&e^{+},\quad e\cdot e=e
\end{eqnarray*}%
with the products for all other pairs being zero, unifying the commutative
tables (\ref{2}), (\ref{3}). It is well known HP (Hudson-Parthasarathy)
table of the vacuum quantum stochastic calculus \cite{1} 
\begin{eqnarray*}
\mathrm{d}\Lambda _{-}\mathrm{d}\Lambda ^{+} &=&I\mathrm{d}t,\quad \mathrm{d}%
\Lambda _{-}\mathrm{d}\Lambda =\mathrm{d}\Lambda _{-}, \\
\mathrm{d}\Lambda \mathrm{d}\Lambda ^{+} &=&\mathrm{d}\Lambda ^{+},\quad 
\mathrm{d}\Lambda \cdot \mathrm{d}\Lambda =\mathrm{d}\Lambda ,
\end{eqnarray*}%
with zero products for all other pairs, for the multiplication of the
canonical number $\mathrm{d}\Lambda $, creation $\mathrm{d}\Lambda ^{+}$,
annihilation $\mathrm{d}\Lambda _{-}$, and preservation $\mathrm{d}\Lambda
_{+}^{-}=I\mathrm{d}t$ differentials in Fock space over the Hilbert space $%
L^{2}\left( \mathbb{R}_{+}\right) $ of square-integrable complex functions $%
f\left( t\right) ,t\in \mathbb{R}_{+}$.

Note that any two-dimensional It\^{o} $\star $-algebra $\mathfrak{a}$ is
commutative as $d_{t}a=0=ad_{t}$ for any other element $a\neq d_{t}$ of the
basis $\left\{ a,d_{t}\right\} $ in $\mathfrak{a}$. Moreover, each such
algebra is either of the Wiener or of the Poisson type, as it is either
second order nilpotent, or contains a unital one-dimensional subalgebra, as
the cases of the subalgebras $\mathfrak{b},\mathfrak{c}$. Other
two-dimensional sub-algebras containing $d_{t}$, are generated by either
Wiener $d_{w}=\bar{\xi}e_{-}+\xi e^{+}$ or Poisson $d_{m}=e+\lambda d_{w}$
element with the special case $d_{m}=e$, corresponding to the only
non-faithful It\^{o} algebra of the Poisson process with zero intensity $%
\lambda ^{2}=0$. However there is only one three dimensional $\star $%
-subalgebra of the four-dimensional HP algebra with $d_{t}$, namely the
noncommutative subalgebra of vacuum Brownian motion, generated by the
creation $e^{+}$ and annihilation $e_{-}$ differentials$.$ Thus our results
on the classification of noncommutative It\^{o} $\star $-algebras will be
nontrivial only in the higher dimensions of $\mathfrak{a}$.

The well known L\'{e}vy-Khinchin classification of the classical noise can
be reformulated in purely algebraic terms as the decomposability of any
commutative It\^{o} algebra into Wiener (Brownian) and Poisson (L\'{e}vy)
orthogonal components. In the general case we shall show that every It\^{o} $%
\star $-algebra is also decomposable into a quantum Brownian, and a quantum L%
\'{e}vy orthogonal components.

Thus classical stochastic calculus developed by It\^{o}, and its quantum
stochastic analog, given by Hudson and Parthasarathy in \cite{1}, was
unified in our $\star $-algebraic approach to the operator integration in
Fock space \cite{3}, in which the classical and quantum calculi become
represented as two extreme commutative and noncommutative cases of a
generalized It\^{o} calculus.

In the next section we remind the definition of the general It\^{o} algebra
given in \cite{3}, and show that every such algebra can be embedded as a $%
\star $-subalgebra into an infinite dimensional vacuum It\^{o} algebra as it
was first proved in \cite{4}.

\section{Representations of It\^{o} $\star $-algebras}

The generalized It\^{o} algebra was defined in \cite{3} as a linear span of
the differentials 
\begin{equation*}
\mathrm{d}\Lambda \left( t,a\right) =\Lambda \left( t+\mathrm{d}t,a\right)
-\Lambda \left( t,a\right) ,\quad a\in \mathfrak{a}
\end{equation*}%
for a family $\left\{ \Lambda \left( a\right) :a\in \mathfrak{a}\right\} $
of operator-valued integrators $\Lambda \left( t,a\right) $ on a pre-Hilbert
space, satisfying for each $t\in \mathbb{R}_{+}$ the $\star $-semigroup
conditions 
\begin{eqnarray}
\mathrm{d}\Lambda \left( t,a\cdot b\right) &=&\mathrm{d}\Lambda \left(
t,a\right) \mathrm{d}\Lambda \left( t,b\right) ,  \label{0.1} \\
\Lambda \left( t,a^{\star }\right) &=&\Lambda \left( t,a\right) ^{\dagger
},\quad \Lambda \left( t,d_{t}\right) =tI,
\end{eqnarray}%
with mean values $\langle \mathrm{d}\Lambda \left( t,a\right) \rangle
=l\left( a\right) \mathrm{d}t$ in a given vector state $\langle \mathrm{%
\cdot }\rangle $, absolutely continuous with respect to $\mathrm{d}t$. Here $%
\Lambda \left( t,a\right) ^{\dagger }$ means the Hermitian conjugation of
the (unbounded) operator $\Lambda \left( t,a\right) $, which is defined on
the pre-Hilbert space for each $t\in \mathbb{R}_{+}$ as the operator $%
\Lambda \left( t,a^{\star }\right) $, 
\begin{equation*}
\mathrm{d}\Lambda \left( t,a\right) \mathrm{d}\Lambda \left( t,b\right) =%
\mathrm{d}\left( \Lambda \left( t,a\right) \Lambda \left( t,b\right) \right)
-\mathrm{d}\Lambda \left( t,a\right) \Lambda \left( t,b\right) -\Lambda
\left( t,a\right) \mathrm{d}\Lambda \left( t,b\right) ,
\end{equation*}%
and $\mathrm{d}t$ is embedded into the family of the operator-valued
differentials as $\mathrm{d}\Lambda \left( t,d_{t}\right) $ with the help of
a special element $d_{t}=d_{t}^{\star }$ of the parametrizing $\star $%
-semigroup $\mathfrak{a}$.

Assuming that the parametrization is exact such that $\mathrm{d}\Lambda
\left( t,a\right) =0\Rightarrow a=0$, where $0=ad_{t}$ for any $a\in 
\mathfrak{a}$, we can always identify $\mathfrak{a}$ with the linear span, 
\begin{equation*}
\sum \lambda _{i}\mathrm{d}\Lambda \left( t,a_{i}\right) =\mathrm{d}\Lambda
\left( t,\sum \lambda _{i}a_{i}\right) ,\quad \forall \lambda _{i}\in 
\mathbb{C},a_{i}\in \mathfrak{a},
\end{equation*}%
and consider it as a complex associative $\star $-algebra, having the death $%
d_{t}\in \mathfrak{a}$, a $\star $-invariant annihilator $\mathfrak{a}\cdot
d_{t}=\left\{ 0\right\} $ corresponding to $\mathrm{d}\Lambda \left( t,%
\mathfrak{a}\right) \mathrm{d}t=\left\{ 0\right\} $. The derivative $l$ of
the differential expectations $a\mapsto l\left( a\right) \mathrm{d}t$ with
respect to the Lebesgue measure $\mathrm{d}t$, called the It\^{o} algebra
state, is a linear positive $\star $-functional 
\begin{equation*}
l:\mathfrak{a}\rightarrow \mathbb{C},\quad l\left( a\cdot a^{\star }\right)
\geq 0,\quad l\left( a^{\star }\right) =\overline{l\left( a\right) },\quad
\forall a\in \mathfrak{a},
\end{equation*}%
normalized as $l\left( d_{t}\right) =1$ correspondingly to the determinism $%
\left\langle I\mathrm{d}t\right\rangle =\mathrm{d}t$ of $\mathrm{d}\Lambda
\left( t,d_{t}\right) $. We shall identify the It\^{o} algebra $\left( 
\mathrm{d}\Lambda \left( \mathfrak{a}\right) ,l\mathrm{d}t\right) $ and the
parametrizing algebra $\left( \mathfrak{a},l\right) $ and assume that it is
faithful in the sense that the $\star $-ideal 
\begin{equation}
\mathfrak{i}=\left\{ b\in \mathfrak{a}:l\left( b\right) =l\left( b\cdot
c\right) =l\left( a\cdot b\right) =l\left( a\cdot b\cdot c\right) =0\quad
\forall a,c\in \mathfrak{a}\right\}   \label{0.2}
\end{equation}%
is trivial, $\mathfrak{i}=\left\{ 0\right\} $, otherwise $\mathfrak{a}$
should be factorized with respect to this ideal. Note that the associativity
of the algebra $\mathfrak{a}$ as well as the possibility of its
noncommutativity is inherited from the associativity and noncommutativity of
the operator product $\Delta \Lambda \left( t,a\right) \Delta \Lambda \left(
t,b\right) $ on the pre-Hilbert space.

Now we can study the representations of the It\^{o} algebra $\left( 
\mathfrak{a},l\right) $. Because any It\^{o} algebra contains the Newton
nilpotent subalgebra $\left( \mathbb{C}d_{t},l\right) $, it has no identity
and cannot be represented by operators in a Euclidean space even if it is
finite-dimensional $\star $-algebra. Thus we have to consider the operator
representations of $\mathfrak{a}$ in a pseudo-Euclidean space, and we shall
find such representations in a Krein space, including the simplest one, a
complex Minkowski space.

Let $\mathbb{K}$ be a complex pseudo-Euclidean space with respect to a
separating indefinite metric $\left( \mathrm{x}|\mathrm{x}\right) $, and $%
\mathrm{k}\mathbf{\in }\mathbb{K}$ be a non-zero vector. We denote by $%
\mathcal{L}\left( \mathbb{K}\right) $ the $\dagger $-algebra of all
operators $\mathrm{A}:\mathbb{K}\rightarrow \mathbb{K}$ with $\mathrm{A}%
^{\dagger }\mathbb{K\subseteq K}$, where $\mathrm{A}^{\dagger }$ is defined
as the kernel of the Hermitian adjoint sesquilinear form $\left( \mathrm{x}|%
\mathrm{A}^{\dagger }\mathrm{x}\right) =\overline{\left( \mathrm{x}|\mathrm{%
Ax}\right) }$. A linear map $\mathrm{i}:\mathfrak{a}\rightarrow \mathcal{L}%
\left( \mathbb{K}\right) $ is a representation of the It\^{o} $\star $%
-algebra $\left( \mathfrak{a},l\right) $ on $\left( \mathbb{K},\mathrm{k}%
\right) $ if 
\begin{equation}
\mathrm{i}\left( a^{\star }\right) =\mathrm{i}\left( a\right) ^{\dagger
},\quad \mathrm{i}\left( a\cdot b\right) =\mathrm{i}\left( a\right) \mathrm{i%
}\left( b\right) ,\quad \left( \mathrm{k}\mathbf{|}\mathrm{i}\left( a\right) 
\mathrm{k}\right) =l\left( a\right) \quad \forall a,b\in \mathfrak{a}\text{.}
\label{0.5}
\end{equation}%
We can always assume that $\left( \mathrm{k}\mathbf{|}\mathrm{k}\right) =0$,
otherwise $\mathrm{k}$ should be replaced by the vector $\mathrm{k}_{+}=%
\mathrm{k}-\frac{1}{2}\left( \mathrm{k}|\mathrm{k}\right) \mathrm{k}_{-}$,
where $\mathrm{k}_{-}=\mathrm{i}\left( d_{t}\right) \mathrm{k}$, with the
same result 
\begin{equation}
\left( \mathrm{k}_{+}|\mathrm{i}\left( a\right) \mathrm{k}_{+}\right)
=l\left( a\right) -\frac{1}{2}\left( \mathrm{k}|\mathrm{k}\right) \left( 
\mathrm{k}\mathbf{|}\mathrm{i}\left( d_{t}a+ad_{t}\emph{\,}-\emph{\,}\frac{1%
}{2}\left( \mathrm{k}|\mathrm{k}\right) d_{t}ad_{t}\right) \mathrm{k}\right)
=l\left( a\right) \text{.}  \label{0.3}
\end{equation}

\begin{proposition}
Every operator representation $\left( \mathbb{K}\text{,}\mathrm{i},\mathrm{k}%
\right) $ of any It\^{o} algebra $\left( \mathfrak{a},l\right) $ is
equivalent to the triangular-matrix representation $\mathbf{i}=\left[ i_{\nu
}^{\mu }\right] _{\nu =-,\circ ,+}^{\mu =-,\circ ,+}$ with $i_{\nu }^{\mu
}\left( a\right) =0$ if $\mu =+$ or $\nu =-$ and $i_{+}^{-}\left( a\right)
=l\left( a\right) $ for all $a\in \mathfrak{a}$. Here $a_{\nu }^{\mu
}=i_{\nu }^{\mu }\left( a\right) $ are linear operators $\mathbb{K}_{\nu
}\rightarrow \mathbb{K}_{\mu }$ on a pseudo-Hilbert (Euclidean if minimal)
space $\mathbb{K}_{\circ }$ and on $\mathbb{K}_{+}=\mathbb{C=K}_{-}$, having
the adjoints $a_{\nu }^{\mu \dagger }:\mathbb{K}_{\mu }\rightarrow \mathbb{K}%
_{\nu }$, which define the pseudo-Hermitian involution $\mathbf{a}\mapsto 
\mathbf{a}^{\dagger }$ by $a_{-\nu }^{\star \mu }=a_{-\mu }^{\nu \dagger }$,
where $-\left( -,\circ ,+\right) =\left( +,\circ ,-\right) $. Moreover, if
the representation is minimal, then $\mathbb{K}_{\circ }$ is a Euclidean
space and $i_{\nu }^{\mu }\left( \text{$d_{t}$}\right) =\delta _{-}^{\mu
}\delta _{\nu }^{+}$.
\end{proposition}

\begin{proof}
In the matrix notation $i_{\nu }^{\mu }\left( a\right) =\mathrm{k}^{\mu }%
\mathrm{i}\left( a\right) \mathrm{k}_{\nu }$, where $\mathrm{k}^{-}=\mathrm{k%
}_{+}^{\dagger }$, $\mathrm{k}^{+}=\mathrm{k}_{-}^{\dagger }$ are defined by 
$\mathrm{k}^{\dagger }\mathrm{x}=\left( \mathrm{k}|\mathrm{x}\right) $ for
all $\mathrm{k}\mathbf{,}\mathrm{x}\in \mathbb{K}$, (\ref{0.3}) can be
written as $i_{+}^{-}\left( a\right) =l\left( a\right) $, and $%
i_{+}^{+}\left( a\right) =0=i_{-}^{-}\left( a\right) $ and $i_{-}^{+}\left(
a\right) =0$ as 
\begin{equation*}
\mathrm{i}\left( a\right) \mathrm{k}_{-}=\mathrm{i}\left( ad_{t}\right) 
\mathrm{k}=0=\mathrm{k}^{\dagger }\mathrm{i}\left( d_{t}a\right) =\mathrm{k}%
^{+}\mathrm{i}\left( a\right) \quad \forall a\in \mathfrak{a}\text{.}
\end{equation*}%
Moreover, due to the pseudo-orthogonality 
\begin{equation*}
\left( \mathrm{x}\mathbf{|}\mathrm{x}\right) =\xi _{-}\xi ^{-}+\left( 
\mathrm{x}_{\circ }|\mathrm{x}_{\circ }\right) +\xi _{+}\xi ^{+}\equiv
\left( \mathbf{x|x}\right) ,
\end{equation*}%
of the decomposition $\mathrm{x}\mathbf{=}\xi ^{-}\mathrm{k}_{-}+\mathrm{x}%
_{\circ }+\xi ^{+}\mathrm{k}_{+}$, where $\xi ^{-}=\mathrm{k}^{-}\mathrm{x}=%
\bar{\xi}_{+}$, $\xi ^{+}=\mathrm{k}^{+}\mathrm{x}=\bar{\xi}_{-}$, $\mathbf{x%
}=\left( \xi _{-},\mathrm{x}_{\circ }^{\dagger },\xi _{+}\right) $, the
representation of the It\^{o} $\star $-algebra $\left( \mathfrak{a},l\right) 
$ is defined by the homomorphism $\mathbf{i}:a\mapsto \left[ i_{\nu }^{\mu
}\left( a\right) \right] $ into the space of triangular block-matrices $%
\mathbf{a}=\left[ a_{\nu }^{\mu }\right] _{\nu =-,\circ ,+}^{\mu =-,\circ
,+} $ with $a_{\nu }^{\mu }=0$ if $\mu =+$ or $\nu =-$.

If the representation is minimal in the sense that $\mathbb{K}\mathbf{=}%
\mathrm{i}\left( \mathfrak{a}\right) \mathrm{k}$, and $\mathrm{k}$ has zero
length, it is pseudo-unitary equivalent to the triangular representation on
the complex Minkowski space $\mathbb{C}\oplus \mathbb{K}_{\circ }\oplus 
\mathbb{C}$, as it can be easily seen in the basis $\mathrm{k}_{+}=\mathrm{k}
$, $\mathrm{k}_{-}=\mathrm{i}\left( d_{t}\right) \mathrm{k}$. Indeed, the
pseudo-orthogonal to the zero length vectors $\mathrm{k}_{-},\mathrm{k}_{+}$
space $\mathbb{K}_{\circ }$ in this case is the complex Euclidean space $%
\mathbb{K}_{\circ }=\left\{ \mathrm{i}\left( a\right) \mathrm{k}:l\left(
a\right) =0\right\} $ as 
\begin{equation*}
\xi ^{+}=\mathrm{k}^{+}\mathrm{i}\left( a\right) \mathrm{k}=0,\quad \xi ^{-}=%
\mathrm{k}^{-}\mathrm{i}\left( a\right) \mathrm{k}=l\left( a\right) \quad
\forall a\in \mathfrak{a},
\end{equation*}%
and $\left( \mathrm{x}_{\circ }|\mathrm{x}_{\circ }\right) =l\left( a^{\star
}\cdot a\right) \geq 0$ for all $\mathrm{x}_{\circ }=\mathrm{i}\left(
a\right) \mathrm{k}-\xi ^{-}\mathrm{k}_{-}=\mathrm{i}\left( a-l\left(
a\right) d_{t}\right) \mathrm{k}$. Moreover, in the minimal representation $%
i_{\circ }^{-}\left( d_{t}\right) =0=i_{+}^{\circ }\left( d_{t}\right) $ and 
$i_{\circ }^{\circ }\left( d_{t}\right) =0$ as 
\begin{equation*}
\mathrm{i}\left( d_{t}\right) \mathrm{x}_{\circ }=\mathrm{i}\left(
d_{t}a\right) \mathrm{k}=0=\mathrm{i}\left( a^{\star }d_{t}\right) \mathrm{k}%
=\mathrm{x}_{\circ }^{\dagger }\mathrm{i}\left( d_{t}\right) \text{ \quad }%
\forall \mathrm{x}_{\circ }\in \mathbb{K}_{\circ }.
\end{equation*}%
Thus, the only nonzero matrix element of $\mathrm{i}\left( d_{t}\right) $ is 
$i_{+}^{-}\left( d_{t}\right) =1.$
\end{proof}

Note that the matrix representation is also defined as the right
representation $\mathbf{x}\mapsto \mathbf{xa}$ on all raw-vectors $\mathbf{x}%
=\left( \xi _{-},\xi ,\xi _{+}\right) $, $\xi \in \mathbb{K}_{\circ
}^{\dagger } $, into the dual space $\mathbb{K}^{*}=\mathbb{C\times K}%
_{\circ }^{*}\mathbb{\times C\supseteq K}^{\dagger }$ with the invariant $%
\mathbb{K}^{\dagger }=\left\{ \mathrm{x}^{\dagger }:\mathrm{x}\in \mathbb{K}%
\right\} $ such that $a_{+}^{-}=\left( \mathrm{k}^{-}\mathbf{a|}\mathrm{k}%
^{-}\right) =l\left( a\right) $, where $\mathrm{k}^{-}=\left( 1,0,0\right) $%
. We shall call the triangular matrix representation on a Minkowski space $%
\mathbb{C}\oplus \mathcal{K}\oplus \mathbb{C}$ canonical if $\mathcal{K}$ is
a minimal pre-Hilbert space. Thus we have proved the second part of the
following

\begin{theorem}
Every It\^{o} $\star $-algebra $\left( \mathfrak{a},l\right) $ can be
canonically realized in a complex Minkowski space. Moreover, every minimal
closed pseudo-Euclidean representation is equivalent to the canonical one in
the Minkowski space.
\end{theorem}

\begin{proof}
Now we construct a faithful canonical operator representation for any It\^{o}
algebra $\left( \mathfrak{a},l\right) $. The functional $l$ defines for each 
$a\in \mathfrak{a}$ the canonical quadruple 
\begin{equation}
a_{\circ }^{\circ }=i\left( a\right) ,\quad a_{+}^{\circ }=k\left( a\right)
,\quad a_{\circ }^{-}=k^{\dagger }\left( a\right) ,\quad a_{+}^{-}=l\left(
a\right) ,  \label{0.6}
\end{equation}
where $i\left( a\right) =i\left( a^{\star }\right) ^{\dagger }$ is the GNS
representation $k\left( ab\right) =i\left( a\right) k\left( b\right) $ of $%
\mathfrak{a}$ in the pre-Hilbert space $\mathbb{K}_{\circ }\ni k\left(
b\right) $, $b\in \mathfrak{a}$ of the Kolmogorov decomposition $l\left(
a\cdot b\right) =k^{\dagger }\left( a\right) k\left( b\right) $, and $%
k^{\dagger }\left( a\right) =k\left( a^{\star }\right) ^{\dagger }$. Such
quadrupole representation $\boldsymbol{i}:a\mapsto \boldsymbol{a}=\left(
a_\nu ^\mu \right) _{\nu =+,\circ }^{\mu =-,\circ }$ of $\mathfrak{a}$ is
multiplicative, $\boldsymbol{i}\left( a\cdot b\right) =\left( a_{\circ }^\mu
b_\nu ^{\circ }\right) _{\nu =+,\circ }^{\mu =-,\circ }$ with respect to the
product given by the convolution of the components $a_\nu $ and $b^\mu $
over the common index values $\mu =\circ =\nu $: 
\begin{eqnarray*}
i\left( a\right) i\left( b\right) &=&i\left( a\cdot b\right) ,\quad
k^{\dagger }\left( a\right) i\left( b\right) =k^{\dagger }\left( a\cdot
b\right) \\
i\left( a\right) k\left( b\right) &=&k\left( a\cdot b\right) ,\quad
k^{\dagger }\left( a\right) k\left( b\right) =l\left( a\cdot b\right) \text{.%
}
\end{eqnarray*}
It is faithful because of the triviality of the ideal (\ref{0.2}). One can
also use the convenience $a_{-}^\mu =0=a_\nu ^{+}$ of the tensor notations (%
\ref{0.6}), extending the quadruples $\boldsymbol{a}=\boldsymbol{i}\left(
a\right) $ to the triangular matrices $\mathbf{a}=\left[ a_\nu ^\mu \right]
_{\nu =-,\circ ,+}^{\mu =-,\circ ,+}$, in which (\ref{0.4}) is simply given
by $\mathbf{i}\left( a\cdot b\right) =\mathbf{ab}$ in terms of the usual
product of the matrices $\mathbf{a}=\mathbf{i}\left( a\right) $ and $\mathbf{%
b}=\mathbf{i}\left( b\right) $. However the involution $a\mapsto a^{\star }$%
, which is given by the Hermitian conjugation $\boldsymbol{i}\left( a^{\star
}\right) =\left( a_{-\mu }^{-\nu \dagger }\right) _{\nu =+,\circ }^{\mu
=-,\circ }$ of the quadruples $\boldsymbol{a}$, where $-(-)=+$, $-\circ
=\circ $, $-(+)=-$, is represented by the adjoint matrix $\mathbf{a}%
^{\dagger }=\mathrm{G}\mathbf{a}^{*}\mathrm{G}$ w.r.t. the pseudo-Euclidean
(complex Minkowski) metric tensor $\mathrm{G}=\left[ \delta _{-\nu }^\mu %
\right] _{\nu =-,\circ ,+}^{\mu =-,\circ ,+}$. Thus, we have constructed the
faithful canonical representation 
\begin{eqnarray}
\mathbf{i}\left( a\right) &=&\left[ 
\begin{array}{lll}
0 & k^{\dagger }\left( a\right) & l\left( a\right) \\ 
0 & i\left( a\right) & k\left( a\right) \\ 
0 & 0 & 0%
\end{array}
\right] ,\quad \mathbf{i}\left( a\cdot b\right) =\mathbf{i}\left( a\right) 
\mathbf{i}\left( b\right) ,  \label{0.7} \\
\mathbf{i}\left( a^{\star }\right) &=&\mathrm{G}\mathbf{i}\left( a\right)
^{*}\mathrm{G}\mathbf{,\quad \quad }\mathrm{G}=\left[ 
\begin{array}{lll}
0 & 0 & 1 \\ 
0 & I & 0 \\ 
1 & 0 & 0%
\end{array}
\right]
\end{eqnarray}
in the Minkowski space $\mathbb{C\oplus }\mathcal{K}\oplus \mathbb{C}$ with $%
\mathcal{K}=k\left( \mathfrak{a}\right) $ and $\mathrm{k}^{-}=\left(
1,0,0\right) $.
\end{proof}

\section{Decomposition of It\^{o} $\star $-algebras}

This was already noted in \cite{3,4} that every (classical or quantum)
stochastic noise described by a process $t\in \mathbb{R}_{+}\mapsto \Lambda
\left( t,a\right) ,a\in \mathfrak{a}$ with independent increments $\mathrm{d}%
\Lambda \left( t,a\right) =\Lambda \left( t+\mathrm{d}t,a\right) -\Lambda
\left( t,a\right) $ forming an It\^{o} $\dagger $-algebra, can be
represented in the Fock space $\mathfrak{F}$ over the space of $\mathcal{K}$%
-valued square-integrable functions on $\mathbb{R}_{+}$ with the vacuum
vector state. This representation is given by $\Lambda \left( t,a\right)
=a_\nu ^\mu \Lambda _\mu ^\nu \left( t\right) $, where 
\begin{equation}
a_\nu ^\mu \Lambda _\mu ^\nu \left( t\right) =a_{\circ }^{\circ }\Lambda
_{\circ }^{\circ }\left( t\right) +a_{+}^{\circ }\Lambda _{\circ }^{+}\left(
t\right) +a_{\circ }^{-}\Lambda _{-}^{\circ }\left( t\right)
+a_{+}^{-}\Lambda _{-}^{+}\left( t\right) ,  \label{0.8}
\end{equation}
is the canonical decomposition of $\Lambda $ into the exchange $\Lambda
_{\circ }^{\circ }$, creation $\Lambda _{\circ }^{+}$, annihilation $\Lambda
_{-}^{\circ }$ and preservation (time) $\Lambda _{-}^{+}=t\mathrm{I}$
operator-valued processes of the vacuum quantum stochastic calculus, having
the mean values $\left\langle \Lambda _\mu ^\nu \left( t\right)
\right\rangle =t\delta _{+}^\nu \delta _\mu ^{-}$, and $a_\nu ^\mu =i_\nu
^\mu \left( a\right) $ are the matrix elements of the canonical
representation associated with the It\^{o} algebra state $l$. If the It\^{o}
algebra $\mathfrak{a}$ is faithful in the sense of the triviality of the
ideal (\ref{0.2}), the constructed canonical representation (\ref{0.7}) is
obviously also faithful. Thus we can identify the faithful algebra $%
\mathfrak{a}$ with the family of quadrupoles $\boldsymbol{a}=\left( i_\nu
^\mu \left( a\right) \right) _{\nu =+,\circ }^{\mu =-,\circ }$, and the
state $l$ on $\mathfrak{a}$ with $l\left( \boldsymbol{a}\right)
=i_{+}^{-}\left( a\right) $, by saying that the It\^{o} algebra is given in
its fundamental representation.

\begin{definition}
Let $\mathcal{K}$ be a pre-Hilbert space, and $\mathfrak{l}\left( \mathcal{K}%
\right) $ be the associated $\star $-algebra of all quadrupoles $A=\left(
a_{\nu }^{\mu }\right) _{\nu =+,\circ }^{\mu =-,\circ }$, where $a_{\nu
}^{\mu }$ are linear operators $\mathbb{K}_{\nu }\rightarrow \mathbb{K}_{\mu
}$ with $\mathbb{K}_{\circ }=\mathcal{K}$, $\mathbb{K}_{+}=\mathbb{C=K}_{-}$%
, having the adjoints $a_{\nu }^{\mu \dagger }:\mathbb{K}_{\mu }\rightarrow 
\mathbb{K}_{\nu }$, with the product and involution 
\begin{equation}
A\cdot B=\left( a_{\circ }^{\mu }b_{\nu }^{\circ }\right) _{\nu =+,\circ
}^{\mu =-,\circ },\quad A^{\ast }=\left( a_{-\mu }^{-\nu \dagger }\right)
_{\nu =+,\circ }^{\mu =-,\circ }\text{.}  \label{0.4}
\end{equation}%
It is an It\^{o} algebra with respect to $l\left( \boldsymbol{a}\right)
=a_{+}^{-}$ and the death $D_{t}=\left( \delta _{-}^{\mu }\delta _{\nu
}^{+}\right) _{\nu =+,\circ }^{\mu =-,\circ }=D_{t}^{\ast }$, $A\cdot
D_{t}=0 $, $\forall A\in \mathfrak{l}\left( \mathcal{K}\right) $, called the
vacuum, or HP (Hudson-Parthasarathy) algebra associated with the space $%
\mathcal{K}$. The fundamental representation of an It\^{o} algebra $\left( 
\mathfrak{a},l\right) $ is given by the constructed canonical homomorphism $%
\boldsymbol{i}:\mathfrak{a}\rightarrow \mathfrak{l}\left( \mathcal{K}\right) 
$ 
\begin{eqnarray*}
\boldsymbol{i}\left( a\right) &=&\left( 
\begin{tabular}{ll}
$l\left( a\right) $ & $k^{\dagger }\left( a\right) $ \\ 
$k\left( a\right) $ & $i\left( a\right) $%
\end{tabular}%
\ \right) ,\,\boldsymbol{i}\left( a^{\star }\right) =\left( 
\begin{tabular}{ll}
$l\left( a^{\star }\right) $ & $k\left( a\right) ^{\dagger }$ \\ 
$k\left( a^{\star }\right) $ & $i\left( a\right) ^{\dagger }$%
\end{tabular}%
\ \right) , \\
\,\boldsymbol{i}\left( a\cdot b\right) &=&\boldsymbol{i}\left( a\right)
\cdot \boldsymbol{i}\left( b\right) ,\quad \boldsymbol{i}\left( a^{\star
}\right) =\boldsymbol{i}\left( a\right) ^{\ast }
\end{eqnarray*}%
into the HP algebra, associated with the space $\mathcal{K}$ of its
canonical representation. An It\^{o} algebra is called vacuum B*-algebra if $%
\mathfrak{n}_{+}^{\perp }=\mathfrak{n}^{-}$, where 
\begin{equation}
\mathfrak{n}_{+}=\left\{ c\in \mathfrak{a}:k\left( c\right) =0\right\}
,\quad \mathfrak{n}^{-}=\left\{ b\in \mathfrak{a}:k^{\dagger }\left(
b\right) =0\right\} ,  \label{1.5}
\end{equation}%
and $\mathfrak{n}_{+}^{\perp }$ is the right (and left) orthogonal
complement to $\mathfrak{n}_{+}$, and it is called thermal algebra if $%
\mathfrak{n}_{+}=\mathbb{C}d_{t}=\mathfrak{n}^{-}$ and the involution $\star 
$ is left (or right) closable on the pre-Hilbert space $\mathcal{D}=%
\mathfrak{a}/\mathbb{C}d_{t}$.
\end{definition}

A subalgebra of $\mathfrak{l}\left( \mathcal{K}\right) $ is a vacuum It\^{o}
algebra iff from orthogonality of $k\left( a\right) $ to all $k\left(
c\right) $ with $k^{\dagger }\left( c\right) =0$ it follows that $k\left(
a\right) =0$. This means the maximality $\mathfrak{n}_{+}=\mathfrak{k}^{-}$
of the left ideal $\mathfrak{n}_{+}=k^{-1}\left( 0\right) \subseteq 
\mathfrak{k}^{-}$, where 
\begin{equation}
\mathfrak{k}^{-}=\left\{ a\in \mathfrak{a}:\left\langle a|b\right\rangle
_{+}=0,\forall b\in \mathfrak{n}^{-}\right\} =\mathfrak{k}_{+}^{\star }
\label{3.6}
\end{equation}
or $\mathfrak{n}^{-}=\mathfrak{k}_{+}$ as the right null ideal $\mathfrak{n}%
^{-}=\mathfrak{n}_{+}^{\star }$ for the map $k^{\dagger }$ in terms of the
right orthogonal complement $\mathfrak{k}_{+}=\mathfrak{n}_{+}^{\perp }$. It
follows from the canonical construction of $\mathcal{K}$ as the quotient
space $\mathfrak{a}/\mathfrak{n}_{+}$. Note that due to the orthogonality of 
$\mathfrak{k}_{+}$ and $\mathfrak{k}^{-}$ in vacuum It\^{o} algebras, the
involution $\star $ is never defined in $\mathfrak{k}_{+}$ or in $\mathfrak{k%
}^{-}$ except on the jointly null ideal $\mathfrak{n}^{-}\cap \mathfrak{n}%
_{+}$.

In the case of thermal It\^{o} algebras the ideals $\mathfrak{n}_{+}$ (and $%
\mathfrak{n}^{-}$) are minimal, and the involution $\star $ is defined into $%
\mathfrak{k}_{+}$ on the whole $\mathfrak{k}_{+}=\mathfrak{a}=\mathfrak{k}%
^{-}$, and thus on the pre-Hilbert space $\mathcal{D}=\mathfrak{a}/\mathbb{C}%
d_{t}$ identified with $\left\{ x=a-l\left( a\right) d_{t}:a\in \mathfrak{a}%
\right\} $, by $x^{\star }=a^{\star }-l\left( a^{\star }\right) d_{t}$ . So
the subalgebra of $\mathfrak{l}\left( \mathcal{K}\right) $ is a thermal
algebra iff the involution is closable on the dense domain $\mathcal{D}%
=k\left( \mathfrak{a}\right) $ of the GNS space $\mathcal{K}$, as it is in
the case of tracial It\^{o} algebras, when the involution is isometric. The
involution $\star $ onto $\mathcal{D}$ has densely defined left and right
adjoints in $\mathcal{D}$ (coinciding with it in the tracial case) iff it is
closable.

We shall call an It\^{o} algebra $\mathfrak{a}$ the Brownian algebra if $%
i\left( \mathfrak{a}\right) =0$, and the L\'{e}vy algebra in the opposite
case, when $i\left( \mathfrak{a}\right) $ is non-degenerated on $k\left( 
\mathfrak{a}\right) $ and thus $i\left( \mathfrak{a}\right) $ has an
identity operator $I\in i\left( \mathfrak{a}\right) $ in the finite
dimensional case. We shall say that an It\^{o} algebra has a quotient
identity $e\in \mathfrak{a}$ if $E=i\left( e\right) =E^{\dagger }$ is the
identity for the operator algebra $\mathcal{A}=i\left( \mathfrak{a}\right) $%
. The following theorem proves that every It\^{o} algebra is an orthogonal
sum of a Brownian algebra and of a L\'{e}vy algebra at least in the finite
dimensional case, as it states the famous L\'{e}vy-Khinchin theorem in the
commutative case. A general infinite dimensional non-commutative version of
the L\'{e}vy-Khinchin decomposition theorem is also true and will be
published elsewhere.

\begin{theorem}
Let $\mathfrak{a}$ be an It\^{o} algebra with a quotient identity. Then it
is an orthogonal sum $\mathfrak{b}+\mathfrak{c}$, $\mathfrak{b}\cdot 
\mathfrak{c}=0$ of a quantum Brownian algebra $\mathfrak{b}$ and a quantum L%
\'{e}vy algebra $\mathfrak{c}$.
\end{theorem}

\begin{proof}
We assume that the quotient algebra $\mathfrak{a}/\mathfrak{n}$ with respect
to the null $\star $-ideal $\mathfrak{n}=\left\{ a\in \mathfrak{a}:i\left(
a\right) =0\right\} $ has an identity, which defines the supporting
ortho-projector $E=i\left( e\right) $ for the operator representation $%
\mathcal{A}=i\left( \mathfrak{a}\right) \simeq \mathfrak{a}/\mathfrak{n}$ on
a pre-Hilbert space $\mathcal{K}$. This means that there exists an element $%
e=e^{\star }\in \mathfrak{a}$ such that $aec=ac$ for all $a,c\in \mathfrak{a}
$, where 
\begin{equation*}
ac=a\cdot c-l\left( a\cdot c\right) d_{t}
\end{equation*}%
is the associative factor-product of the $\star $-algebra $\mathfrak{a}/%
\mathbb{C}d_{t}\simeq \left\{ a\in \mathfrak{a}:l\left( a\right) =0\right\} $
for the zero mean elements $a-l\left( a\right) d_{t}$. Ideed, it is so in
the canonical representation (\ref{0.7}) as $i\left( ae\right) =i\left(
a\right) =i\left( ea\right) $ for all $a\in \mathfrak{a}$ because $i\left(
d_{t}\right) =0$ and $i\left( e\right) =E$, 
\begin{equation*}
k\left( aec\right) =i\left( ae\right) k\left( c\right) =k\left( ac\right)
,\quad k^{\dagger }\left( aec\right) =k^{\dagger }\left( c\right) i\left(
ea\right) =k^{\dagger }\left( ac\right) ,
\end{equation*}%
and $l\left( aec\right) =0=l\left( ac\right) $. We assume that $e$ is an
idempotent, otherwise it should be replaced by $e^{2}$.

We can easily then define the required orthogonal decomposition $\mathfrak{a}%
=\mathfrak{b}+\mathfrak{c}$ by 
\begin{equation*}
a=b+c,\quad c=ae+ea-eae.
\end{equation*}%
Here $b$ is an element of the quantum Brownian algebra $\mathfrak{b}=\left\{
b\in \mathfrak{a}:be=0=eb\right\} \subseteq \mathfrak{n}$ which is
orthogonal to the subalgebra $\mathfrak{aa}$ as 
\begin{equation*}
b\cdot a=bea+l(b\cdot a)d_{t}=l\left( b\cdot a\right) d_{t},\quad a\cdot
b=aeb+l\left( a\cdot b\right) d_{t}=l\left( a\cdot b\right) d_{t}
\end{equation*}%
for all $b\in \mathfrak{b}$, and hence 
\begin{equation*}
b\cdot aa^{\star }=ba\cdot a^{\star }=0=a^{\star }\cdot ab=a^{\star }a\cdot b
\end{equation*}%
for all $a\in \mathfrak{a}$. And $c$ is an element of a quantum L\'{e}vy
algebra $\mathfrak{c}$, the closure of $\mathfrak{aa}$ in $\mathfrak{a}$
which coincides with all the algebraic combinations $c$ as $c^{\star
}c=a^{\star }a$ for all $a\in \mathfrak{a}$. Thus $a=b+c$, $b\cdot c=0$ for
all $a\in \mathfrak{a}$, where $b\in \mathfrak{b}$ is in a Brownian algebra
with the fundamental representation $\mathfrak{b}\subseteq \mathbb{C}\oplus 
\mathcal{G}_{+}\oplus \mathcal{G}^{-}$, where $\mathcal{G}_{+}=Pk\left( 
\mathfrak{a}\right) $, $P=I-E$, $\mathcal{G}^{-}=k^{\dagger }\left( 
\mathfrak{a}\right) P$, and $c\in \mathfrak{c}$ is in a L\'{e}vy algebra,
having the fundamental representation $\mathfrak{c}\subseteq \mathbb{C\oplus 
\mathcal{E}}_{+}\oplus \mathcal{E}^{-}\mathbb{\oplus }\mathcal{A}$ with
non-degenerated operator algebra $\mathcal{A}=i\left( \mathfrak{a}\right) $,
left and right represented on $\mathcal{E}_{+}=Ek\left( \mathfrak{a}\right) $
and $\mathcal{E}^{-}=k^{\dagger }\left( \mathfrak{a}\right) E$.
\end{proof}

\section{Vacuum and Thermal It\^{o} algebras}

Here we consider the two extreme cases of It\^{o} algebras as sub-algebras
of the vacuum algebra $\mathfrak{l}\left( \mathcal{K}\right) $ associated
with a pre-Hilbert space $\mathcal{K}$. The first case corresponds to a pure
state $l$ on $\mathfrak{a}$ as it is in the case of a quantum noise of zero
temperature, and the second case corresponds to a completely mixed $l$ as in
the case of a quantum noise of a finite temperature.

\subsection{Vacuum noise $\star $-algebra}

Let $\mathcal{K}$ be a pre-Hilbert space of ket-vectors $\zeta $ with scalar
product $\left( \zeta |\zeta \right) $ and $\mathcal{A}\subseteq \mathcal{L}%
\left( \mathcal{K}\right) $ be a $\star $-algebra, represented on $\mathcal{K%
}$ by the operators $\mathcal{A}\ni A:\zeta \mapsto A\zeta $ with the
adjoints $\left( A^{\dagger }\zeta |\xi \right) =\left( \zeta |A\xi \right) $%
, $A^{\dagger }\mathcal{K}\subseteq \mathcal{K}$. We denote by $\mathcal{K}%
^{\dagger }$ the dual space of bra-vectors $\eta =\zeta ^{\dagger }$, $\zeta
\in \mathcal{K}$ with the scalar product $\left( \eta |\xi ^{\dagger
}\right) =\eta \xi =\left( \eta ^{\dagger }|\xi \right) $, $\xi \in \mathcal{%
K}$ given by inverting anti-linear isomorphism $\mathcal{K}^{\dagger }\ni
\eta \mapsto \eta ^{\dagger }\in \mathcal{K}$, and the dual representation
of $\mathcal{A}$ as the right representation $A^{\prime }:\eta \mapsto \eta
A $, $\eta \in \mathcal{K}^{\dagger }$, given by $\left( \eta A\right) \zeta
=\eta \left( A\zeta \right) $ such that $\left( \eta A^{\dagger }|\eta
\right) =\left( \eta |\eta A\right) $ on $\mathcal{K}^{\dagger }$. Then the
direct sum $\mathcal{K}\oplus \mathcal{K}^{\dagger }$ of $\xi =\zeta \oplus
\eta $ becomes a two-sided $\mathcal{A}$-module 
\begin{equation}
A\left( \zeta \oplus \eta \right) =A\zeta ,\quad \left( \zeta \oplus \eta
\right) A=\eta A,\quad \forall \zeta \in \mathcal{K},\eta \in \mathcal{K}%
^{\dagger },  \label{1.1}
\end{equation}
with the flip-involution $\xi ^{\star }=\eta ^{\dagger }\oplus \zeta
^{\dagger }$ and two scalar products 
\begin{equation}
\left\langle \zeta \oplus \eta ^{\prime }|\zeta ^{\prime }\oplus \eta
\right\rangle _{+}=\left( \zeta |\zeta ^{\prime }\right) ,\quad \left\langle
\zeta \oplus \eta ^{\prime }|\zeta ^{\prime }\oplus \eta \right\rangle
^{-}=\left( \eta ^{\prime }|\eta \right) .  \label{1.2}
\end{equation}

The space $\mathfrak{a}=\mathbb{C}\oplus \mathcal{K}\oplus \mathcal{K}%
^{\dagger }\oplus \mathcal{A}$ of triples $a=\left( \alpha ,\xi ,A\right) $
becomes an It\^{o} $\star $-algebra with respect to the non-commutative
product 
\begin{equation}
a^{\star }\cdot a=\left( \left\langle \xi |\xi \right\rangle _{+},\xi
^{\star }A+A^{\dagger }\xi ,A^{\dagger }A\right) ,\quad a\cdot a^{\star
}=\left( \left\langle \xi |\xi \right\rangle ^{-},A\xi ^{\star }+\xi
A^{\dagger },AA^{\dagger }\right) ,  \label{1.3}
\end{equation}%
where $\left( \alpha ,\xi ,A\right) ^{\star }=\left( \bar{\alpha},\xi
^{\star },A^{\dagger }\right) $, with death $d_{t}=\left( 1,0,0\right) $ and 
$l\left( \alpha ,\xi ,A\right) =\alpha $. Obviously $a^{\star }\cdot a\neq
a\cdot a^{\star }$ if $\left\Vert \xi \right\Vert _{+}=\left\Vert \zeta
\right\Vert \neq $ $\left\Vert \eta \right\Vert =\left\Vert \xi \right\Vert
^{-}$ even if the operator algebra $\mathcal{A}$ is commutative, $A^{\dagger
}A=AA^{\dagger }$.

We shall call such It\^{o} algebra the vacuum algebra as $l\left( a^{\star
}\cdot a\right) =0$ for any $a\in \mathfrak{a}$ with $\xi \in \mathcal{K}%
^{\dagger }$ (the Hudson-Parthasarathy algebra $\mathfrak{a}=\mathfrak{l}%
\left( \mathcal{K}\right) $ if $\mathcal{A}=\mathcal{L}\left( \mathcal{K}%
\right) $). Every It\^{o} algebra is a subalgebra $\mathfrak{a}\subseteq 
\mathfrak{l}\left( \mathcal{K}\right) $ of the HP algebra $\mathfrak{l}%
\left( \mathcal{K}\right) $ for a pre-Hilbert space $\mathcal{K}$ with the
operator factor-algebra $\mathcal{A}$ represented on $\mathcal{K}$.

If the algebra $\mathcal{A}$ is completely degenerated on $\mathcal{K}$, $%
\mathcal{A}=\left\{ 0\right\} $, the It\^{o} algebra $\mathfrak{a}$ is
nilpotent of second order, and contains only two-dimensional subalgebras of
Wiener type $\mathfrak{b}=\mathbb{C}\oplus \mathbb{C}\oplus \left\{
0\right\} $ generated by an $a=\left( \alpha ,\zeta \oplus \eta ,0\right) $
with $\left\| \zeta \right\| =\left\| \eta \right\| $. Every It\^{o}
subalgebra $\mathfrak{b}\subseteq \mathfrak{a}$ of the HP algebra $\mathfrak{%
a}=\mathfrak{l}\left( \mathcal{K}\right) $ is called the It\^{o} algebra of
a vacuum Brownian motion if it is defined by a $\star $-invariant direct sum 
$\mathcal{G}\oplus \mathcal{G}^{\dagger }$ given by a subspace $\mathcal{G}%
\subseteq \mathcal{K}$ and $\mathcal{A}=\left\{ 0\right\} $.

In the case $I\in \mathcal{A}$ the algebra $\mathcal{A}$ is not degenerated
and contains also the vacuum Poisson subalgebra $\mathbb{C}\oplus \left\{
0\right\} \oplus \mathbb{C}I$ of the total quantum number on $\mathcal{K}$,
and other Poisson two-dimensional subalgebras, generated by $a=\left( \alpha
,\zeta \oplus \eta ,I\right) $ with $\eta =e^{i\theta }\zeta ^{\dagger }$.
We shall call a closed It\^{o} subalgebra $\mathfrak{c}\subseteq \mathfrak{a}
$ of the HP algebra $\mathfrak{a}=\mathfrak{l}\left( \mathcal{K}\right) $
the algebra of a vacuum L\'{e}vy motion if it is given by a direct sum $%
\mathcal{E}\oplus \mathcal{E}^{\dagger }$ and a $\star $-subalgebra $%
\mathcal{A}\subseteq \mathcal{L}\left( \mathcal{K}\right) $ non-degenerated
on the subspace $\mathcal{E}\subseteq \mathcal{K}$.

Our decomposition theorem for the vacuum algebras can be reformulated as
follows

\begin{theorem}
Every vacuum algebra $\mathfrak{a}$ having the unital factor-algebra $%
\mathcal{A} $ can be decomposed into an orthogonal sum $\mathfrak{a}=%
\mathfrak{b}+\mathfrak{c}$, $\mathfrak{b}\cdot \mathfrak{c}=\left\{
0\right\} $ of the Brownian vacuum algebra $\mathfrak{b}$ and the L\'{e}vy
vacuum algebra $\mathfrak{c}$.
\end{theorem}

\begin{proof}
This decomposition is uniquely defined for all $a=\left( \alpha ,\xi
,A\right) $ by $a=\alpha d_{t}+b+c$, with $b=\left( 0,\eta ,0\right) $, $%
c=\left( 0,\zeta ,A\right) $, $\eta =P\xi \oplus \xi P\in \mathcal{G}$, $%
\zeta =\xi -\eta \in \mathcal{E}$, where $P=I-E=P^{\dagger }$ is the maximal
projector in $\mathcal{K}$, for which $\mathcal{A}P=\left\{ 0\right\} $, $%
\mathcal{G}=P\mathcal{K}$, and $\mathcal{E}=\mathcal{G}^{\perp }$ is the
range of the identity orthoprojector $E\in \mathcal{A}$.
\end{proof}

\subsection{Thermal noise $\star $-algebra}

Let $\mathcal{D}$ be a left $\star $-algebra \cite{5} with respect to a
Hilbert norm $\left\| \xi \right\| _{+}=0\Rightarrow \xi =0$, and thus a
right pre-Hilbert $\star $-algebra with respect to $\left\| \xi \right\|
^{-}=\left\| \xi ^{\star }\right\| _{+}$. This means that $\mathcal{D}$ is a
complex Euclidean space with left (right) multiplications $C:\zeta \mapsto
\xi \zeta $ ($C^{\prime }:\eta \mapsto \eta \xi $) w.r.t. $\left\| \cdot
\right\| _{+}$ (w.r.t. $\left\| \cdot \right\| ^{-}$) of the elements $\zeta
,\eta \in \mathcal{D}$ respectively, defined by an associative product in $%
\mathcal{D}$, and the involution $\mathcal{D}\ni \xi \mapsto \xi ^{\star
}\in \mathcal{D}$ such that 
\begin{equation}
\left\langle \eta \zeta ^{\star }|\xi \right\rangle ^{-}=\left\langle \eta
|\xi \zeta \right\rangle ^{-},\quad \left\langle \eta ^{\star }\zeta |\xi
\right\rangle _{+}=\left\langle \zeta |\eta \xi \right\rangle _{+}\quad
\forall \xi ,\zeta ,\eta \in \mathcal{D},  \label{2.1}
\end{equation}
were $\left\langle \eta |\xi ^{\star }\right\rangle ^{-}=\left\langle \eta
^{\star }|\xi \right\rangle _{+}$ is the right scalar product. The
involution is assumed to have the adjoints 
\begin{equation}
\left\langle \eta |\xi ^{\star }\right\rangle ^{-}=\left\langle \xi |\eta
^{\sharp }\right\rangle ^{-},\quad \left\langle \zeta |\xi ^{\star
}\right\rangle _{+}=\left\langle \xi |\zeta ^{\flat }\right\rangle _{+}\quad
\forall \eta \in \mathcal{D}^{-},\zeta \in \mathcal{D}_{+},  \label{2.2}
\end{equation}
where $\mathcal{D}_{+}=\mathcal{D}_{+}^{\flat }$ is a dense domain for the
left adjoint involution $\zeta \mapsto \zeta ^{\flat }$, $\zeta ^{\flat
\flat }=\zeta $, and $\mathcal{D}^{-}=\mathcal{D}_{+}^{\star }$ is the dense
domain for the right adjoint involution $\eta \mapsto \eta ^{\sharp }$, $%
\left( \eta ^{\sharp }\eta \right) ^{\sharp }=\eta ^{\sharp }\eta $ such
that $\zeta ^{\flat \star }=\zeta ^{\star \natural }$, $\eta ^{\sharp \star
}=\eta ^{\star \flat }$.

Note that we do not require the sub-space $\mathcal{DD}\subseteq \mathcal{D}$
of all products $\left\{ \eta \zeta :\eta ,\zeta \in \mathcal{D}\right\} $
to be dense in $\mathcal{D}$ w.r.t. any of two Hilbert norms on $\mathcal{D}$%
. Hence the operator factor-algebra $\mathcal{C}=\left\{ C:\mathcal{D}\ni
\zeta \mapsto \xi \zeta |\xi \in \mathcal{D}\right\} $ w.r.t. the left
scalar product, which is also represented on the $\mathcal{D}\ni \eta $
equipped with $\left\langle \cdot |\cdot \right\rangle ^{-}$ by the right
multiplications $\eta C=\eta \xi $, $\xi \in \mathcal{D}$, can be
degenerated on $\mathcal{D}$.

Thus the direct sum $\mathfrak{a}=\mathbb{C}\oplus \mathcal{D}$ of pairs $%
a=\left( \alpha ,\xi \right) $ becomes an It\^{o} $\star $-algebra with the
product 
\begin{equation}
a^{\star }a=\left( \left\langle \xi |\xi \right\rangle _{+},\xi ^{\star }\xi
\right) ,\quad aa^{\star }=\left( \left\langle \xi |\xi \right\rangle
^{-},\xi \xi ^{\star }\right) ,  \label{2.4}
\end{equation}%
where $\left( \alpha ,\xi \right) ^{\star }=\left( \bar{\alpha},\xi ^{\star
}\right) $, with death $d_{t}=\left( 1,0\right) $ and $l\left( \alpha ,\xi
\right) =\alpha $. Obviously $a^{\star }a\neq aa^{\star }$ if the involution 
$a\mapsto a^{\star }$ is not isometric w.r.t. any of two Hilbert norms even
if the algebra $\mathcal{D}$ is commutative.

We shall call such It\^{o} algebra the thermal It\^{o} algebra as $l\left(
a^{\star }a\right) =\left\| \xi \right\| _{+}^2\neq 0$ for any $a\in 
\mathfrak{a} $ with $\xi \neq 0$. If $\zeta \eta =0$ for all $\zeta ,\eta
\in \mathcal{D}$, it is the It\^{o} algebra of a thermal Brownian motion. A
thermal subalgebra $\mathfrak{b}\subseteq \mathfrak{a}$ with such trivial
product is given by any involutive pre-Hilbert $\star $-invariant two-normed
subspace $\mathcal{G}\subseteq \mathcal{D}$. We shall call such Brownian
algebra $\mathfrak{b}=\mathbb{C\oplus }\mathcal{G}$ the quantum (if $\left\|
\cdot \right\| _{+}\neq $ $\left\| \cdot \right\| ^{-}$) Wiener algebra
associated with the space $\mathcal{G}$.

In the opposite case, if $\mathcal{DD}=\left\{ \zeta \eta :\zeta ,\eta \in 
\mathcal{D}\right\} $ is dense in $\mathcal{D}$, it has non-degenerated
operator representation $\mathcal{C}$ on $\mathcal{D}$. Any involutive
sub-algebra $\mathcal{E}\subseteq \mathcal{D}$ which is non-degenerated on $%
\mathcal{E}$ defines an It\^{o} algebra $\mathfrak{c}=\mathbb{C}\oplus 
\mathcal{E}$ of thermal L\'{e}vy motion. We shall call such It\^{o} algebra
the quantum (if $\mathcal{E}$ is non-commutative) Poisson algebra.

Our decomposition theorem for the thermal algebras can be reformulated as
follows

\begin{theorem}
Every thermal It\^{o} algebra $\mathfrak{a}$ having the unital subalgebra $%
\mathcal{DD}$ is an orthogonal sum $\mathfrak{a}=\mathfrak{b}+\mathfrak{c}$, 
$\mathfrak{bc}=\left\{ 0\right\} $ of the Wiener algebra $\mathfrak{b}$ and
the Poisson algebra $\mathfrak{c}$.
\end{theorem}

\begin{proof}
The orthogonal decomposition $a=\alpha d_{t}+b+c$ for all $a=\left( \alpha
,\xi \right) \in \mathfrak{a}$, uniquely given by the decomposition $\xi
=\eta +\zeta $ w.r.t. any of two scalar products in $\mathcal{D}$, where $%
\eta =P\xi =\xi P$ is the orthogonal projection onto $\mathcal{G}\perp 
\mathcal{DD}$ w.r.t. any of two Hilbert norms, and $\zeta =\xi -\eta $ .

Indeed, if $\xi \in \mathcal{D}$ is left orthogonal to $\mathcal{DD}$, then
it is also right orthogonal to $\mathcal{DD}$ and vice versa: 
\begin{eqnarray*}
\left\langle \eta \zeta ^{\star }|\xi \right\rangle ^{-} &=&\left\langle
\zeta \eta ^{\star }|\xi ^{\star }\right\rangle _{+}=\left\langle \xi |\eta
^{\sharp \star }\zeta ^{\flat }\right\rangle _{+}=0,\quad \forall \eta \in 
\mathcal{D}^{-},\zeta \in \mathcal{D}_{+}, \\
\left\langle \eta ^{\star }\zeta |\xi \right\rangle _{+} &=&\left\langle
\zeta ^{\star }\eta |\xi ^{\star }\right\rangle ^{-}=\left\langle \xi |\eta
^{\sharp }\zeta ^{\flat \star }\right\rangle ^{-}=0,\quad \forall \eta \in 
\mathcal{D}^{-},\zeta \in \mathcal{D}_{+}.
\end{eqnarray*}
From these and (\ref{2.1}) equations it follows that $\eta \xi =0=\xi \zeta $
for all $\zeta ,\eta \in \mathcal{D}$ if $\xi $ is (right or left)
orthogonal to $\mathcal{DD}$, and so $i\left( \xi \right) =0$ for such $\xi $
and vice versa. Thus the orthogonal subspace $\mathcal{G}=\left\{ \xi \in 
\mathcal{D}:i\left( \xi \right) =0\right\} $ is the range of the
orthoprojector $P=I-E$, where $E$ is the identity orthoprojector of $%
\mathcal{C}$, representing the unity $\varepsilon =\varepsilon ^{\star }$ of 
$\mathcal{DD}$ such that $\varepsilon \xi =\zeta =\xi \varepsilon \in 
\mathcal{D}$ is in $\mathcal{E}=\mathcal{DD}$ and $\eta =\xi -\zeta \in 
\mathcal{G}$.
\end{proof}

\begin{example}
The commutative multiplication table $\mathrm{d}b_i\mathrm{d}\bar{b}%
_k=\delta _k^i\mathrm{d}t$ for the complex It\^{o} differentials $\mathrm{d}%
b_k=\mathrm{d}\bar{b}_{-k}$, $k=0,\pm 1,\ldots ,\pm K$ of the complex
amplitudes 
\begin{equation*}
b_k\left( t\right) =\sum_{n=1}^Ne^{jk\theta _n}w_n\left( t\right) ,\quad 
\mathrm{d}w_i\left( t\right) \mathrm{d}w_n\left( t\right) =\frac 1N\delta
_n^i\mathrm{d}t
\end{equation*}
for $N$ independent Wiener processes $w_n,n=1,\ldots ,N$ with $N\geq
2K+1,\theta _n=\frac{2\pi }Nn-\pi $ can be generalized in the following way.

Let $\rho _{k}>0$, $k=0,\pm 1,\ldots ,\pm K$ be a self-inverse family of
spectral eigen-values $\rho _{-k}=\rho _{k}^{-1}$. The generalized
multiplication table $d_{i}\cdot d_{k}^{\star }=\rho _{k}\delta
_{k}^{i}d_{t} $ for abstract infinitesimals $d_{k}=d_{-k}^{\star }$, $k\in 
\mathbb{Z}$ is obviously non-commutative for all $k$ with $\rho _{k}\neq 1$.
The $\star $-semigroup $\left\{ 0,\text{$d_{t}$},d_{k}:\left\vert
k\right\vert \leq K\right\} $ generates a $2\left( K+1\right) $-dimensional
It\^{o} algebra $\left( \mathfrak{b},l\right) $ of a quantum Wiener periodic
motion on $\left[ -\pi ,\pi \right] $ as it is the second order nilpotent
algebra $\mathfrak{a} $ of $a=\alpha d_{t}+b$, $y=\sum \eta ^{k}d_{k}$ with $%
y^{\star }=\sum \eta _{k}d_{k}^{\star }$ and $l\left( y\right) =0$ for all $%
\eta ^{k}=\bar{\eta}_{k}\in \mathbb{C}$. It is a Brownian algebra with
closed involution on the complex space $\mathcal{D}$ of all $\eta $ given by
all complex sequences $\eta =\left( \eta _{k}\right) _{\left\vert
k\right\vert \leq K}$. The operator representation of $\mathfrak{a}$ in Fock
space is defined by the forward differentials of $\Lambda \left( t,a\right)
=\alpha t\mathrm{I}+\sum \eta ^{k}\hat{w}_{k}\left( t\right) $, where $\hat{w%
}_{k}\left( t\right) =\hat{v}_{k}^{-}\left( t\right) +\hat{v}_{k}^{+}\left(
t\right) $, $\,$%
\begin{equation*}
\hat{v}_{k}^{-}\left( t\right) =\left( \frac{\rho _{-k}}{N}\right) ^{\frac{1%
}{2}}\sum_{n=1}^{N}e^{jk\theta _{n}}\Lambda _{-}^{n}\left( t\right) ,\quad 
\hat{v}_{k}^{+}\left( t\right) =\left( \frac{\rho _{k}}{N}\right) ^{\frac{1}{%
2}}\sum_{n=1}^{N}e^{jk\theta _{n}}\Lambda _{n}^{+}\left( t\right)
\end{equation*}%
are given by the annihilation and creation measures in Fock space over
square integrable functions on $\mathbb{R}_{+}\times \left[ -\pi ,\pi \right]
$ with the standard multiplication table 
\begin{equation*}
\mathrm{d}\Lambda _{-}^{n}\left( t\right) \mathrm{d}\Lambda _{m}^{+}\left(
t\right) =\delta _{m}^{n}I\mathrm{d}t,\quad \mathrm{d}\Lambda _{m}^{+}\left(
t\right) \mathrm{d}\Lambda _{-}^{n}\left( t\right) =0.
\end{equation*}
\end{example}

\begin{example}
The commutative multiplication table $\mathrm{d}c_i\mathrm{d}\bar{c}%
_k=\delta _k^i\mathrm{d}t+\mathrm{d}c_{i-k}$ for the complex It\^{o}
differentials $\mathrm{d}c_k=\mathrm{d}\bar{c}_{-k}$, $k=0,\pm 1,\ldots ,\pm
K$ of the complex amplitudes $c_k\left( t\right) =\sum_{n=1}^Ne^{jk\theta
_n}m_n\left( t\right) $, $\theta _n=\theta _n=\frac{2\pi }Nn-\pi $, given by 
$N\geq 2K+1$ compensated Poisson processes $m_n\left( t\right) $ with 
\begin{equation*}
\mathrm{d}m_i\left( t\right) \mathrm{d}m_n\left( t\right) =\frac 1N\delta
_n^i\mathrm{d}t+\mathrm{d}m_n\left( t\right) \delta _n^i,\quad i,n=1,\ldots
,N,
\end{equation*}
can be generalized in the following way.

Let $G$ be a finite group, and $G\mathbb{\ni }g\mapsto \lambda _{g}\in 
\mathbb{C}$ be a positive-definite function, $\lambda _{g^{-1}}=\bar{\lambda}%
_{g}$, which is self-inverse in the convolutional sense $\left[ \bar{\lambda}%
\ast \lambda \right] _{g}=\sum_{h}\bar{\lambda}_{gh^{-1}}\lambda _{h}=\delta
_{g}^{1}$. The generalized multiplication table for abstract infinitesimals 
\begin{equation*}
d_{g}=d_{-g}^{\star },\quad d_{g}\cdot d_{h}^{\star }=\lambda _{gh^{-1}}%
\text{$d_{t}$}+d_{gh^{-1}},\quad g,h\in G
\end{equation*}%
is obviously associative and commutative if $G$ is Abelian as in the above
case, but it is non-commutative for non Abelian $G$ even if $\lambda
_{k}=\delta _{k}^{1}$ as in the above case. The $\star $-semigroup $\left\{
0,\text{$d_{t}$},d_{g}:g\in G\right\} $ generates finite dimensional It\^{o}
algebra $\left( \mathfrak{a},l\right) $ of a quantum compensated Poisson
motion on the spectrum $\Omega $ of the group $G$ as $\mathfrak{a}$ is the
sum of $\mathbb{C}d_{t}$ and the unital group algebra $\mathcal{D}$ of $%
z=\sum \zeta ^{g}d_{g}$ with involution $z^{\star }=\sum \zeta
_{g}d_{g}^{\star }$ and $l\left( z\right) =0$ for all $\zeta ^{g}=\bar{\zeta}%
_{g}\in \mathbb{C}$. It is a L\'{e}vy algebra with closed involution on the
complex space $\mathcal{D}$ of all complex sequences $\zeta =\left( \zeta
_{g}\right) _{g\in G}$ with 
\begin{equation*}
\left\Vert \zeta \right\Vert ^{-}=\left( \sum \left( \bar{\zeta}\ast \tilde{%
\zeta}\right) _{g}^{2}\lambda _{g}\right) ^{1/2}=\left\Vert \zeta ^{\star
}\right\Vert _{+},
\end{equation*}%
where $\tilde{\zeta}=\left( \zeta _{g^{-1}}\right) _{g\in G}$. The operator
representation of $\mathfrak{a}$ in Fock space is defined over the Hilbert
space of square integrable function on $\mathbb{R}_{+}$ with values in the
direct sum $\mathcal{K}=\oplus _{n\in \hat{G}}\mathcal{K}\left( n\right) 
\mathrm{d}_{n}$ of finite dimensional Euclidean spaces $\mathcal{K}\left(
n\right) $ for unitary irreducible representations $U_{g}\left( n\right) $
of spectrum $\hat{G}$ of the group $G$ with the Plansherel measure $\mathrm{d%
}_{n},n\in \hat{G}$. It is given by the forward differentials of $\Lambda
\left( t,a\right) =\alpha t\mathrm{I}+\sum \zeta ^{g}\hat{m}_{g}\left(
t\right) $, where $\,$%
\begin{equation*}
\hat{m}_{g}\left( t\right) =\sum_{n\in N}\mathrm{Tr}_{\mathcal{K}\left(
n\right) }\left\{ U_{g}\left( n\right) \left[ \rho _{n}^{1/2}\Lambda
_{-}^{n}\left( t\right) +\Lambda _{n}^{+}\left( t\right) \rho
_{n}^{1/2}+\Lambda _{n}^{n}\left( t\right) \right] \right\} ,
\end{equation*}%
with the standard annihilation, creation and exchange operators in this Fock
space, and a family $\left( \rho _{n}\right) _{n\in \hat{G}}$ of positive
operators in $\mathcal{K}\left( n\right) $ with the traces $\mathrm{Tr}\rho
_{n}$, defining the spectral decomposition 
\begin{equation*}
\lambda _{g}=\sum_{n\in \hat{G}}\mathrm{Tr}\left[ \rho _{n}U_{g}\left(
n\right) \right] \mathrm{d}_{n}.
\end{equation*}
\end{example}


\begin{thebibliography}{9}
\bibitem{Ito51} It\^{o}, K. On a Formula Concerning Stochastic
Differentials. Nagoya Math . J., \textbf{3}, pp. 55-65, 1951.

\bibitem{Grd91} Gardiner, C.W. Quantum Noise. Springer-Verlag, 1991.

\bibitem{1} Hudson, R.\thinspace L.\ and Parthasarathy, K.\thinspace R.
Quantum It\^{o}'s Formula and Stochastic Evolution. Comm.\thinspace
Math.\thinspace Phys., \textbf{93}, pp.\thinspace 301--323, 1984.

\bibitem{3} Belavkin, V.\thinspace P. Chaotic States and Stochastic
Integration in Quantum Systems. Russian Math.\thinspace Survey, \textbf{47},
(1), pp.\thinspace 47--106, 1992.

\bibitem{4} Belavkin V.P. Kernel Representations of $\star $-semigroups
Associated with Infinitely Divisible States. Quantum Probability and Related
Topics, Vol VII, pp 31--50, 1992

\bibitem{5} Takesaki, M. J. Functional Analysis \textbf{9}, p 306, 1972.
\end{thebibliography}
\end{document}